\documentclass[twocolumn]{autart}    

\usepackage{amsmath} 
\usepackage{amssymb}

\usepackage{graphics} 
\usepackage{subfig}
\usepackage{epsfig} 
\graphicspath{ {images/} }
\usepackage{color,soul} 

\usepackage{times} 
\usepackage{cite}

\usepackage{url}

\usepackage{bbm}
\usepackage{amsthm}

\theoremstyle{remark}
\newtheorem{remark}{Remark}
\newtheorem{theorem}{Theorem}

\newtheorem{lemma}{Lemma}

\usepackage{enumerate}
\begin{document}

\begin{frontmatter}

\title{Exactly Optimal Bayesian Quickest Change Detection for Hidden Markov Models}

\thanks[footnoteinfo]{%
J.~Ford acknowledges continued support from the Queensland University of Technology (QUT) through the Centre for Robotics.
Corresponding author J.~J.~Ford.}

\author[qut]{Jason J.~Ford}\ead{j2.ford@qut.edu.au},    
\author[uq]{Jasmin James}\ead{jasmin.martin@uq.edu.au}  and  
\author[anu]{Timothy L.~Molloy}\ead{Timothy.Molloy@anu.edu.au},     

\address[qut]{School of Electrical Engineering and Robotics, University of Technology (QUT), Brisbane, QLD 4000, Australia} 
\address[uq]{School of Mechanical \& Mining Engineering, University of Queensland (UQ), Brisbane, QLD 4000, Australia} 
\address[anu]{School of Engineering, Australian National University (ANU),  Acton, ACT 2601, Australia}  

\begin{abstract}
This paper considers the quickest detection problem for hidden Markov models (HMMs) in a
Bayesian setting. We construct 
an augmented HMM representation of the problem 
 that allows the
application of a dynamic programming approach to prove that Shiryaev's rule is an (exact) optimal solution. 
This augmented representation highlights the problem's fundamental information structure and suggests possible relaxations to more exotic change event priors not appearing in the literature. Finally, this augmented representation  
allows us to present an efficient computational method for implementing the optimal solution. 
 
\end{abstract}

\end{frontmatter}
 
\section{Introduction}
Quickest change detection (QCD) problems are concerned
with the quickest (on-line) detection of a change in the statistical properties of an observed process.
Such problems naturally arise in a wide variety of applications including quality control \cite{nikiforov},  
target detection \cite{jamesCST} and fault detection \cite{nikiforov,Hwang},
in which we desire an alert of a possible change event quickly (as soon as possible) subject to a constraint on the occurrence of false alarms. This paper is concerned with QCD for the case of hidden Markov model processes.

There are various formulations for QCD problems that differ by assumptions on the point of change and optimality criteria. 
Early theoretical formulations for quickest change detection were developed by Shiryaev under the assumption that the change point is a random variable with a known geometric distribution and the observations are independent and identically distributed (i.i.d.) 
\cite{shiryaev2007optimal}.
These early theoretical formulations are classified as Bayesian formulations since they assume that the change point is a random variable.
Shiryaev established an optimal (stopping) rule which compares the posterior probability of a change with a threshold. 
Shiryaev's formulation has since been extended to encompass non-geometrically distributed change-times 
\cite{Tartakovsky05,krishnamurthy2016}
 and dependent data (i.e., non-i.i.d. observations) \cite{Dayanik,yakir,tartakovsky2005,Fuh}. 

Despite various (generalised) Bayesian QCD formulations appearing in the literature, establishing optimal detection rules for dependent data 
has remained a challenging problem.
In \cite{yakir} some progress was made by showing that an optimal rule for QCD for Markov chain process is a Bayes rule which depends on the current state of the chain. 
Further, it was recently established for QCD of a statistically periodic process that a stopping rule based on a periodic sequence of thresholds is exactly optimal \cite{Banerjee21}.
In \cite{Dayanik} an $\epsilon$-optimal approach to the related joint HMM QCD \emph{and} identification problem  
was investigated which provide some insights into the connections between hidden Markov models (HMMs) and Bayesian QCD.
The difficulty of finding (exactly) optimal detection rules for non-i.i.d. observations has led to the development of weaker asymptotic optimality results that hold as the probability of false alarms vanishes.
Hence, the strongest results for Bayesian QCD for dependent process are \cite{Tartakovsky05} which show Shiryaev's rule is asymptotically optimal in the general non-i.i.d. case and in \cite[Eq. (21)]{Fuh} for a generalised HMMs case  (generalised in the sense of having measurements conditional on both the current Markov state and the previous measurement).
 
In this paper we develop exact (non-asymptotic)
optimal solutions to Bayesian QCD
for the standard class of HMMs whose measurements are conditional on the current Markov state 
and not being conditional on the previous measurement as in \cite{Fuh} when considering a delay penalty that is independent of the Markov chain process (unlike the chain process dependent cost considered in \cite{Banerjee21}). Although we slightly restrict the problem compared to \cite{Fuh}, we are the first to establish exact optimality results in an HMM setting.
For this purpose, we show this Bayesian QCD for HMM problem can be recast into an augmented HMM representation 
which enables us to exploit standard dynamic programming tools to establish that Shiryaev's rule is exactly optimal  
(further, we note this augmented representation suggests possible relaxations to 
more exotic priors not appearing in the literature).
Specifically, the paper's contributions are:
 \begin{itemize}
     \item Establishing  Shiryaev's rule is an (exactly) optimal rule for Bayesian QCD for HMMs (noting that existing results
     hold only in the asymptotic regime).
     \item  Presenting an efficient recursion for calculating the posterior information required to apply Shiryaev's rule.
\end{itemize}

\section{Bayesian HMM QCD}\label{sec:HMMQCD}
 This section presents the  
Bayesian HMM QCD problem.  
\subsection{State and Observation Process}
Let  us first define two finite state spaces $S_b \triangleq \{e^b_1,\dots, e^b_{N_b} \}$ and
$S_a \triangleq \{e^a_1,\dots, e^a_{N_a}\}$  
where $e^b_i \in \mathbb{R}^{N_b}$ and $e^a_i \in \mathbb{R}^{N_a }$ are indicator vectors with $1$ in the $i$th element and zeros elsewhere, and $N_b \ge 1$ and $N_a \ge 1$ are the  
HMM order before and after the change event, respectively.

For $k \geq 0$, we consider a process $X_k$ which is able to randomly transition between states in the space of the current stage (within $S_b$ or $S_a$) or able to transition to a state in the space of the next stage (from $S_b$ to $S_a$).
We assume $X_k$ starts in the first stage in the sense $X_0 \in S_b$ and has probability $p(X_0)$. 
For $k <\nu$, $X_k \in S_b$  
can be modelled a first-order time-homogeneous Markov chain described by the transition probabilities $A_b^{i,j} \triangleq  P( X_{k+1}=e^b_i | X_{k+1} \in S_b,  X_{k}=e^b_j  )$ for $1 \le i,j \le N_b$.
At some unknown time $k=\nu$, where we assume $\nu\ge1$,  
 $X_k$ transitions between stages in the sense $X_{\nu -1} \in S_b$ and $X_{\nu} \in S_a$ 
according to state change probabilities $A_\nu^{i,j} \triangleq  P( X_{k+1}=e^a_i | X_{k+1}  
\in S_a,  X_{k}=e^b_j  )$ for  $1 \le i \le N_a$ and $1 \le j \le N_b$.  
For $k > \nu$, $X_k \in S_a$  
can be modelled as a first-order time-homogeneous Markov chain described by the transition probabilities $A_a^{i,j} \triangleq  P( X_{k+1}=e^a_i |  X_{k+1} \in S_a,  X_{k}=e^a_j  )$, for $1 \le i,j \le N_a$.

Finally, for each $k > 0$, $X_k$ is  observed through a stochastic process $y_k \in \mathcal{Y}$ generated by conditional observation densities
$b_b(y_k,i)\triangleq P(y_k|X_k=e^b_i)$ for $1 \le i \le N_b$ and $k < \nu$ and $b_a(y_k,i) \triangleq P(y_k|X_k=e^a_i)$ for $1 \le i \le N_a$ and $k \ge \nu$.  Let $X_{[0,k]} \triangleq  \{X_0, \ldots, X_k\}$ and $y_{[1,k]} \triangleq  \{y_1, \ldots, y_k\}$ be short hand for state and measurement sequences.

\subsection{Probability Measure Space}
Before we formally state our Bayesian HMM QCD problem,  let us first introduce a probability measure space.  Let $\mathcal{F}_k=\sigma(X_{[0,k]},y_{[1,k]})$ denote the filtration generated by $X_{[0,k]},y_{[1,k]}$. We will assume the existence of a probability space $(\Omega,\mathcal{F},P_\nu)$ where 
we consider the set $\Omega$ consisting of all infinite sequences  
$\omega \triangleq (X_{[0,\infty]}; y_{[1,\infty]})$.  
Since $\Omega$ is separable and a complete metric space it can be endowed with a Borel $\sigma$-algebra $\mathcal{F}=\cup_{k=1}^\infty \mathcal{F}_k$ with the convention that $\mathcal{F}_0=\{0,\Omega\}$, and $P_\nu$ is the probability measure constructed using Kolmogorov's extension on the joint probability density function of the state and observations $p_\nu (X_{[0,k]},y_{[1,k]})$. For $k<\nu$  we can model the joint probability density function of the state and observations by
\begin{equation*}
\begin{split}
      p_\nu(X_{[0,k]}&,y_{[1,k]}) \triangleq\\
    & \left(\Pi_{\ell=1}^{k}
 b_b(y_{\ell},\zeta(X_\ell)) A_b^{\zeta(X_\ell),\zeta(X_{\ell-1})} \right)  p(X_{0}) 
\end{split}
\end{equation*}
where $\zeta(e_i) \triangleq i$ returns the index of the non-zero element of an indicator vector $e^b_i$ or $e^a_i$. For $k\ge \nu$  we can model the joint probability density function of the state and observations by
\[
p_\nu (X_{[0,k]},y_{[1,k]}) \triangleq 
p_b (X_{[0,\nu]},y_{[1,\nu]}) p_a(X_{[\nu+1,k]},y_{[\nu+1,k]} | X_\nu)
\]
where the joint probability of state and observations up to the change time is given by
\begin{equation*}
\begin{split}
 &    p_b (X_{[0,\nu]},y_{[1,\nu]}) \triangleq  b_a(y_{\nu},\zeta(X_{\nu}))   A_\nu^{\zeta(X_\nu),\zeta(X_{\nu-1})}\\
  &
 \times \left( \Pi_{\ell=1}^{\nu-1}
 b_b(y_{\ell},\zeta(X_\ell)) A_b^{\zeta(X_\ell),\zeta(X_{\ell-1})} \right)  p(X_{0})
\end{split}
\end{equation*}
and the joint probability of state and observations after change time is given by 
\begin{equation*}
\begin{split}
p_a &(X_{[\nu+1,k]},y_{[\nu+1,k]}| X_\nu) \triangleq  \\
&
\Pi_{\ell=\nu+1}^{k}
 b_a(y_{\ell},\zeta(X_{\ell})) A_a^{\zeta(X_{\ell}),\zeta(X_{\ell-1})}.
\end{split}
\end{equation*}
and we define $p_a(X_{[\nu+1,k]},y_{[\nu+1,k]}| X_\nu) \triangleq 1$ if $k<\nu+1$.
We will let $E_\nu$ denote expectation under $P_\nu$.

\subsection{Change Time Prior}
Under the Bayesian QCD formulation we consider the change time $\nu \ge 1$ to be an unknown random variable with prior distribution  $\pi_k \triangleq P(\nu=k)$ for $k\ge 1$ for $G \in \mathcal{F}$ 
This allows us to construct a new averaged measure $P_\pi(G)=\sum_{k=1}^{\infty} \pi_k(G) P_k(G)$ for all $G \in \mathcal{F}$ and we let $E_\pi$ denote the corresponding expectation operation. 
In this presentation, the geometric prior $\pi_k=(1-\rho)^{k-1} \rho$ with $\rho \in ( 0,1)$ as introduced by Shiryaev  
\cite{shiryaev2007optimal}.
 
\subsection{Bayesian QCD for HMMs: Shriyaev Formulation}

The classic formulation of Bayesian QCD seeks to find a stopping time $\tau \ge 1$ with respect to the filtration generated by $y_{[1,k]}$ (having knowledge of $p(X_0)$) that solves the following constrained optimisation problem
\begin{equation}
  \inf_{\tau \in T(\alpha)} E_\pi [ (\tau - \nu)^+ ]
\label{equ:concost}  
\end{equation}
where $(\tau-\nu)^+ \triangleq \max(0, \tau-\nu)$
and $T(\alpha) \triangleq \{ \tau : P_\pi (\tau < \nu) \le \alpha \}$ denotes the set of stopping times satisfying a given probability of false alarm constraint $\alpha \in (0,1-\rho)$ (noting we are only interested in $\alpha < 1-\rho$ as $\alpha\ge 1-\rho$ has the trivial optimal solution of $\tau=0$).

Alternatively, the relaxed Bayes formulation of the QCD problem seeks to find a stopping time $\tau \ge 1$ with respect to the filtration generated by $y_{[1,k]}$ (having knowledge of $p(X_0)$) that solves the unconstrained  
optimisation problem 
\begin{equation}
 \inf_{\tau \in T(1)} J(\tau), \;
    J(\tau) \triangleq c E_\pi \left[(\tau-\nu)^+ \right] + P_\pi(\tau < \nu) 
    \label{eqn:cost} 
 \end{equation}
  for some 
  $c>0$
  which is the penalty on each time step that alert is not declared after $\nu$, and  
  $(\tau -\nu)^+ \triangleq \max(0,\tau -\nu)$.
As recently established in  
\cite{Dey2022,Banerjee2012},
\eqref{eqn:cost} is a Lagrangian relaxation of \eqref{equ:concost}, and thus it can be seen that if $c$ can be found such that the solution to \eqref{eqn:cost} achieves the probability of false alarm constraint with equality, then the solution to \eqref{eqn:cost} is also the solution to \eqref{equ:concost}.
 
This work extends the $\epsilon$-optimal and asymptotic optimality results for Bayesian HMM QCD in \cite{Dayanik,Fuh} to an exact optimality result. Note here the change identification aspects of  \cite{Dayanik} are not considered and 
a standard HMM is considered having measurements $P(y_{k}|X_{[0,k]},y_{[1,k-1]})=P(y_{k}|X_k)$ not being conditional on the previous measurement, rather than the generalised HMM  considered in \cite[Eq. (21)]{Fuh} with measurements
$P(y_{k}|X_{[0,k]},y_{[1,k-1]})=P(y_{k}|X_k,y_{k-1})$ with potential conditioning on the previous measurement.

\section{Main Result} \label{sec:AbstractConstruction}
In this section we present a generalised augmented construction of a Bayesian HMM change detection problem, which we will use to establish our main optimality result for Bayesian  HMM QCD.  

\subsection{An Augmented HMM Representation}\label{sec:augmentAndMode}

We   define a new augmented state process  $Z_k \in S$ where  $S \triangleq \{e_1,\dots, e_{N} \}$ where $e_i \in \mathbb{R}^{N}$ (are indicator vectors with 1 in the $i$th element and zero elsewhere) and $N=N_b +N_a$.  This augmented state process combines the information of $X_k$ and $\nu$
as follows.  
  For $k<\nu$, $Z_k \in S$ is defined as
\begin{equation*}
Z_k \triangleq \begin{bmatrix}
       X_k \\
      \mathbf{0}_a
\end{bmatrix},
\end{equation*}
and for $k\geq \nu$ as
\begin{equation*}
Z_k \triangleq \begin{bmatrix}
      \mathbf{0}_b \\
      X_k
\end{bmatrix}.
\end{equation*}
where $\mathbf{0}_b$ and $\mathbf{0}_a$ are the zero vectors of size $N_b$ and $N_a$, respectively.

\begin{lemma}\label{lem:hmm}
The augmented process $Z_k$ is a first-order time-homogeneous Markov chain with transition probabilities  $A^{i,j} \triangleq  P_\pi( Z_{k+1}=e_i |  Z_{k}=e_j)$ that can be written as 
\[
   A =\left[
\begin{array}{cc}
(1-\rho) A_b  & \mathbf{0}_{b \times a}  \\
\rho  A_\nu  &  A_a
\end{array} 
\right]\\
\]
where  
$\mathbf{0}_{b \times a}$ is a $N_b \times N_a$ matrix of all zeros. Moreover with measurement matrix 
$B^{j,j}(y_k) \triangleq  P_\pi( y_k |  Z_{k}=e_j)$ of
\begin{align*}
    B(y_k) = 
    \text{diag}&(b_b(y_k,1), \dots, \\
    &b_b(y_k,{N_b}),    b_a(y_k,1),\dots ,b_a(y_k,{N_a}))
\end{align*}
then   ($Z_k, y_k)$ are the   state  and observation processes of a  hidden Markov model with transition matrix $A$ and measurement matrix $B$.

\end{lemma}
\begin{proof}
We establish this result by considering $A$ to be a block matrix
made from the 4 types of different transitions between sets $S_b$ and $S_a$.
First, looking at pre-change self-transition (type $X_{k} \in S_b$ and $X_{k+1} \in S_b$) 
we note from Bayes rule, for all $i,j \in \{1,\dots, N_b\}$, we can write
\begin{align*}
P_\pi( X_{k+1}=e^b_i |  X_{k}=e^b_j) 
= \\
 P_\pi( X_{k+1}&=e^b_i| X_{k+1} \in S_b, X_{k}=e^b_j) \\ 
&\times P_\pi( X_{k+1} \in S_b |  X_{k}=e^b_j) 
\end{align*}
where from previous definitions we have  
$P_\pi( X_{k+1}=e^b_i |  X_{k}=e^b_j) = (1-\rho) A_b^{i,j}$, leading to the matrix block $(1-\rho) A_b$.  The other blocks can be determined in a similar manner.
 
To establish the measurement matrix we first define the function $\eta({e}_i) \triangleq (m,n)$ which takes the indicator vector of the augmented process ${e}_i$.   From the definition of $Z_k$ note that $P(y_k|Z_k={e}_i) = b^m(y_k,n)$ where $(m,n)=\eta({e}_i)$, and hence the second lemma result follows.
\end{proof}
This HMM representation lets us derive our optimal rule
which can be efficiently calculated.  
\begin{remark}
Although not considered here, this augmented HMM representation is flexible enough to consider state dependent change priors more general than typically considered in the literature (e.g. when the change event has dependence on the current value of the pre-change state).
\end{remark}
 
\subsection{Optimal Quickest Detection Rule}\label{sec:QCD}
We now present our main result establishing that an optimal rule for Bayesian QCD of HMMs is a simple threshold test.

To facilitate analysis, let $\hat{Z}_k^i\triangleq  P_\pi (Z_k = e_i| y_{[1,k]})$  denote the posterior probabilities of being in each of the states of $Z_k$    with initial conditions $\hat{Z}_0$, where $\hat{Z}_0^i = P(Z_0 = e^b_i)$ for $i\in \{1, \dots, N_b\}$ and $\hat{Z}_0^i = 0$ elsewhere. 
We can define the operation $M({Z})\triangleq \sum_{i=1}^{N_b} {Z}^i$ 
and no change posterior $\hat{M}_k^1 \triangleq  M(\hat{Z}_k)$.  

We can now introduce an auxiliary QCD cost function corresponding to an auxiliary QCD problem that starts at some general time $k\ge0$ as follows 
\begin{equation}
\bar{J}(\tau,k,\hat{Z}_k) \triangleq 
E_\pi \left.  \left[ c \sum_{\ell=k}^{\tau -1}    (1-M( {Z}_\ell) )+ M( {Z}_\tau )\right|   \hat{Z}_k  \right]
\label{equ:costz}
\end{equation}
and note we recover our standard cost function when $k=0$ in the sense that $J(\tau)=\bar{J}(\tau,0,\hat{Z}_0)$. It is useful to define a value function $V(\hat{Z})\triangleq \min_{\tau} \bar{J}(\tau,1,\hat{Z})$
in terms of the first time instant that a change could occur.  

 We now  present a preliminary lemma result needed 
 for the main theorem. 

\begin{lemma} \label{lem:valuesame}
Let $M_1 \in [0,1]$ be a possible value of $\hat{M}^1_k$ and let $\mathcal{S}(M_1) \triangleq \{\hat{Z}  : M(\hat{Z} )=M_1\}$ represent all the possible value of $\hat{Z} $ which lead to $\hat{M}^1_k=M_1$. Then the value function $V(\hat{Z})$  
has the same value for all $\hat{Z}\in \mathcal{S}(M_1)$. 
\end{lemma}
\begin{proof}
Consider any $k\ge 0$ and any  $\ell \in \{k,k+1,\ldots \}$. Let $A^{(\ell-k)}$ denote the $(\ell-k)$ power of $A$, it then follows that 
\begin{align*}
E_\pi[{M}(Z_\ell)| \hat{Z}_k] & = M(E_\pi[Z_\ell| \hat{Z}_k] )  
\\
&
= M( A^{(\ell-k)}\hat{Z}_k) \\
& =\sum_{i=1}^{N_b} \sum_{j=1}^{N_b} ((1-\rho) ^{(\ell -k )}A_b^{(\ell-k)})^{i,j}\hat{Z}_k \\
&= (1-\rho) ^{(\ell-k)} \sum_{j=1}^{N_b} \hat{Z}_k  \\
&= (1-\rho) ^{(\ell-k)} M(\hat{Z}_k) 
\end{align*}
where the first step follows as $M(\cdot)$ is a linear operation, the second step follows due expectation properties of Markov chains \cite[Ch. 2]{elliott1995}, the third step follows from the definition of matrix operations and the structure of $A$, the fourth step follows because rows of transition probabilities matrices sum to one, and the final step follows from the definition of $M(\cdot)$.

We are now able to establish the lemma claim.  At any step, a stopping rule $\tau$ can either stop or continue. At some $k\ge 0$ we can consider the auxiliary QCD cost \eqref{equ:costz} to understand this choice and write that
\begin{equation*}
\bar{J}(\tau,k,\hat{Z}_k) = \left\{ \begin{array}{ll}
\mbox{if stop} & E_\pi [  M(Z_k) | \hat{Z}_k]    \\
  \mbox{otherwise}& E_\pi [  c(1-M(Z_k)) | \hat{Z}_k]  \\
 & \ + c  E_\pi [ \sum_{\ell=k+1}^{\tau-1} (1- M(Z_\ell)) \\
 & \ \  +  M(Z_\tau) | \hat{Z}_k]   
\end{array} \right.
\end{equation*}
Using above result that $E_\pi[{M}(Z_\ell)| \hat{Z}_k] =(1-\rho) ^{(\ell-k)} M(\hat{Z}_k) $, then $\bar{J}(\tau,k,\hat{Z}_k) $ can be written as
\begin{align*}
&\bar{J}(\tau,k,\hat{Z}_k) = \\
& \ \ \ \quad \left\{ \begin{array}{ll}
\mbox{if stop} &  M(\hat{Z}_k)   \\ 
\mbox{otherwise} &  c(1-M(\hat{Z}_k))  + E_\pi [ \sum_{\ell=k+1}^{\tau-1}  c  | \hat{Z}_k]    \\
& +  E_\pi [ \sum_{\ell=k+1}^{\tau-1} c (1- \rho)^{(\ell-k)}    | \hat{Z}_k]    M(\hat{Z}_k) \\
& +   E_\pi [  (1- \rho)^{(\tau -k )}    | \hat{Z}_k]   M(\hat{Z}_k)
\end{array} \right.
\end{align*}

Hence $\bar{J}(\tau,k,\hat{Z}_k) $ only depends on $c$, $\rho$, the value of $M(\hat{Z}_k)$ and $E_\pi[\cdot| \hat{Z}_k]$ terms whose value depends only on policy choice. Given the above form, the cost of stopping being $M(\hat{Z}_k)$ implies that if the optimal policy is to stop at some $\hat{Z}_k$, with $M(\hat{Z}_k)=M_1$,
then all other elements of 
$\hat{Z} \in \mathcal{S}(M_1)$ 
have the same valued $M(\hat{Z})=M_1$ terms appearing in their stop \& continue cost terms and hence must also have that the optimal policy is to stop (conversely, if the optimal policy was to continue for some $\hat{Z}_k$, with $M(\hat{Z}_k)=M_1$, then there cannot be a different $\hat{Z} \in \mathcal{S}(M_1)$ such that the optimal policy is to stop, otherwise as $\hat{Z}_k$ has the same cost choices and it would have also been optimal policy to stop at $\hat{Z}_k$).
Hence, the different values of $\hat{Z}  \in \mathcal{S}(M_1)$ must have the same minimising action. 
Setting $k=1$ and using definition of value function gives that $V(\hat{Z})$ has the same value for all $\hat{Z}\in \mathcal{S}(M_1)$ and hence the lemma claim.
\end{proof}

Our main optimality result for Bayesian HMM QCD follows. 
\begin{theorem}\label{thm:main}
For the cost criterion \eqref{eqn:cost},
the optimal HMM QCD rule with stopping time $\tau^{*}$, is a threshold check of no change posterior against threshold 
$h \ge 0$ given by
\begin{equation}
\label{eqn:qcdRule}
\tau^*=\inf \{ k \ge 1: \hat{M}^1_k \leq h \}.
\end{equation}
\end{theorem} 
\begin{proof}
Approach here is similar to used in \cite{jamesCST} for {\it i.i.d.} processes. 
The value function $V(\hat{Z})$
corresponding to our cost criterion \eqref{eqn:cost} can described by the recursion (Bellman's Equation)   
\cite[pg. 258]{krishnamurthy2016} and \cite[Section 3.4]{Bert05}:
\begin{equation*}\label{eqn:val}
\begin{split}
V(\hat{Z} ) = \min &\bigg\{c  (1-M(\hat{Z} ) )\\
+&  
E_\pi \left[ V\left(\hat{Z}^+(\hat{Z} ,y_{k+1})\right) \bigg| \hat{Z}   \right] ,  M(\hat{Z} )  \bigg\},
\end{split}
\end{equation*}
where $\hat{Z}^+(\hat{Z},y) = \langle \underline{1},B(y) A \hat{Z} \rangle^{-1} B(y)A \hat{Z} $, and $B(y) = \text{diag}(b_b(y,1), \dots, b_b(y,{N_b}), b_a(y,1),\dots ,b_a(y,{N_a}))$ and $\underline{1}$ is the  
 vector of all ones. Moreover, for $\hat{Z}$ such that $ M(\hat{Z} ) \le V(\hat{Z})$ then the optimal action is to stop,  
 otherwise the optimal action is to continue.
 
Let $\mathcal{R}_S \triangleq \{\hat{Z}  : V(\hat{Z} )=M(\hat{Z} )\}$ 
denote the optimal stopping set that we are seeking. Using a similar approach to \cite[sec. 12.2.2]{krishnamurthy2016},  and noting that the cost is linear here, then according to  \cite[Theorem 7.4.2]{krishnamurthy2016}, 
$V(\hat{Z})$ are 
concave in $\hat{Z}$. We can then use \cite[Thm. 12.2.1]{krishnamurthy2016} and \cite[Page 164]{Bert05} to show that the stopping set $\mathcal{R}_S$  is convex.

If 
$\hat{Z}=e^a_i$, 
for any $i\in \{1,\ldots, N_a \}$,  
then
$M(e^a_i)=0$ gives
 \begin{equation*}
V(\hat{Z})  = \min \left\{ c +  
E_\pi \left[ V\left(\hat{Z}^+(\hat{Z} ,y)\right) \bigg|  \hat{Z}  \right],   0  \right\}.
\end{equation*}
Since $V\left(\hat{Z}^+(\hat{Z} ,y)\right) $ is non-negative  then
$V(\hat{Z} )  = 0$, which shows 
$e^a_i$
belongs to the stopping set.

Then note that Lemma  
\ref{lem:valuesame} provides
that $V(\hat{Z}_k)$ has the same value for all $\hat{Z}_k \in \mathcal{S}(M_1)$ which implies  
the convex stopping set $\mathcal{R}_S $ 
is equivalent to a convex stopping interval on $M(\hat{Z}_k)$ of the form $0 \le d \le h \le 1$, for some $h \in R$ and $d \in R$. 

Since $V\left(\hat{Z}^+(\hat{Z} ,y)\right) $ is non-negative  then
$V(\hat{Z} )  = 0$, which shows $M(\hat{Z})=0$ belongs to the stopping set, thus $d=0$ and $\mathcal{R}_S$ is an interval of the form  $[0,h]$. We can express the optimal  stopping time as the first time that the stopping set $\mathcal{R}_S$ is reached giving our theorem result.
 \end{proof}
\section{Example}
In this section, we illustrate our (exactly) optimal rule \eqref{eqn:qcdRule} in an example involving a two state HMM which changes to a three state HMM with a geometric prior $\rho = 0.0005$, for $T=10 000$ timesteps. Consider a situation with the following transition probabilities:
\begin{align*}
   & A_b = \begin{bmatrix} 0.99 & 0.01 \\
    0.01 & 0.99
    \end{bmatrix},
      A_a = \begin{bmatrix} 0.90 & 0.05 &0.05 \\
    0.05 & 0.90 & 0.05 \\
      0.05 & 0.05 & 0.90  
    \end{bmatrix} \mbox{and} \\
    &   A_\nu = \begin{bmatrix} 0.999 & 0.999   \\
      0.0005& 0.0005 \\
     0.0005 & 0.0005 
    \end{bmatrix}.
\end{align*}

The observation measurements $y_k$ are i.i.d. with marginal probability densities (before) $b_b(y,1) = \psi(y-1)$ and $b_b(y,2) = \psi(y-1.2)$ and (after) $b_a(y,1) = b_b(y,1)$,  $b_a(y,2) = b_b(y,2)$and $b_a(y,1) = \psi(y-2.5)$ where $\psi(\cdot)$ is a zero mean Gaussian probability density function with variance $\sigma^2 = 1$.

We note that at time $k\ge1$ the test statistic $\hat{M}_k^1=M(\hat{Z}_k )$ can be efficiently calculated 
via the  HMM filter for $\hat{Z}_k$   \cite{elliott1995}
\begin{equation}\label{eqn:CME}
\hat{Z}_k = N_{k} B(y_k) A\hat{Z}_{k-1} 
\end{equation}
with scalar normalisation  $N_{k} \triangleq \langle \underline{1}, B(y_k)A\hat{Z}_{k-1} \rangle^{-1}$ where $\underline{1}$ is the $N \times 1$ vector of all ones.

\begin{figure}
\begin{center}
\includegraphics[scale=0.5,trim={1.0cm 0.0cm 0.0cm 0.0cm}]{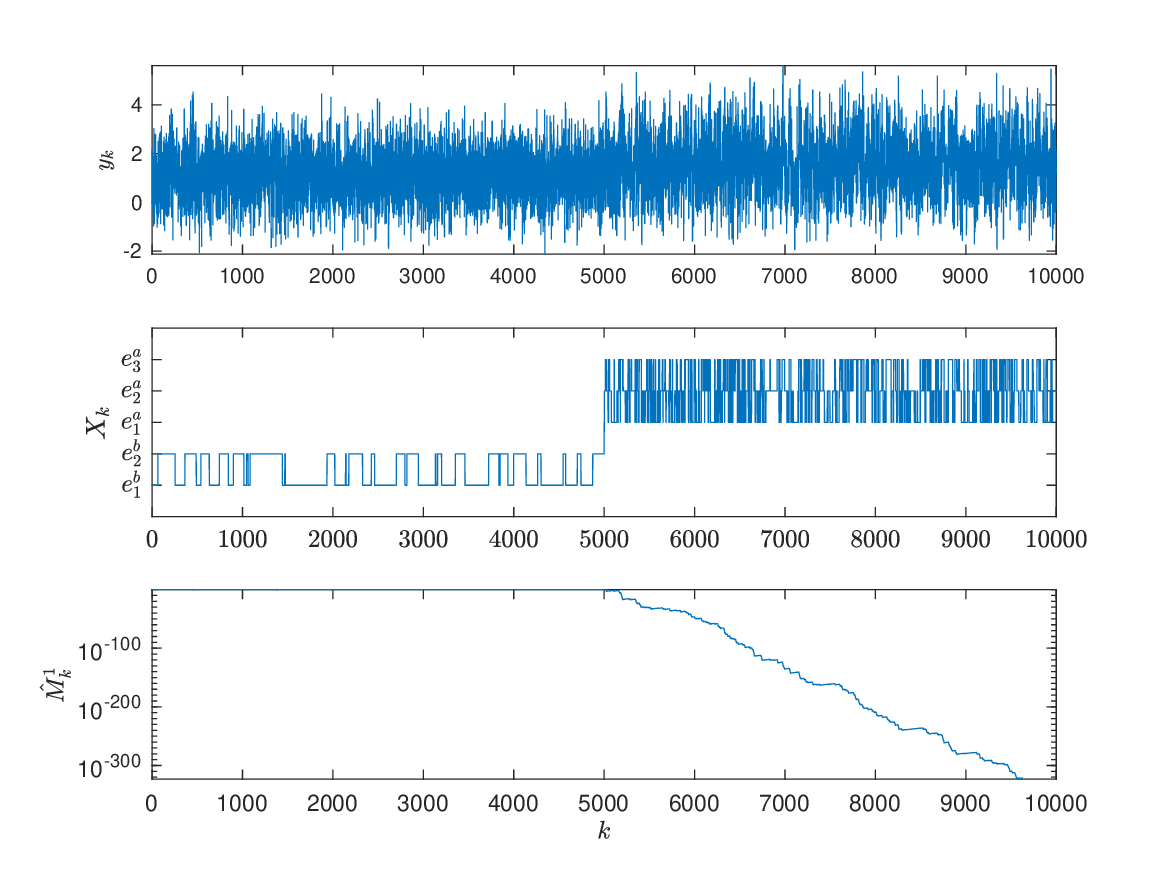}
  \caption{Example: Change event at $k=5001$. From top to bottom: the measurements $y_k$,  the underlying state process $X_k$, and 
  posterior $\hat{M}^1_k$.
}
\label{fig:sim}
\end{center}
\end{figure}

An illustrative example with a change event at $\nu=5001$ is shown 
in Figure \ref{fig:sim} with the 
measurement sequence $y_k$ (top), state process $X_k$  (middle) and posterior $M^1_k$ (bottom).
The posterior suggests no difficulty in selecting $h$ in \eqref{eqn:qcdRule} for reliable detection.

\section{Discussion}
Theorem \ref{thm:main} characterises the nature of the optimal rule for HMM QCD and is the first to establish an exact optimality result for Bayesian HMM QCD (previous results in 
\cite{Tartakovsky05, Fuh, Banerjee21} 
are limited to the asymptotic setting, admittedly allowing slightly generalised problem settings).

\bibliography{ref}

\end{document}